\documentstyle[12pt,emulateapj]{article}
 
\input{epsf}

\journalid{}{}
\articleid{}{}
\begin{document}
\def\lax    {\ifmmode{_<\atop^{\sim}}\else{${_<\atop^{\sim}}$}\fi}
\def\gax    {\ifmmode{_>\atop^{\sim}}\else{${_>\atop^{\sim}}$}\fi}
\def\gtorder{\mathrel{\raise.3ex\hbox{$>$}\mkern-14mu
             \lower0.6ex\hbox{$\sim$}}}
\def\ltorder{\mathrel{\raise.3ex\hbox{$<$}\mkern-14mu
             \lower0.6ex\hbox{$\sim$}}}
 
\long\def\***#1{{\sc #1}}
 
\title{Millennium Year X-ray Transients in Andromeda Galaxy}

\author{Sergey P. Trudolyubov\altaffilmark{1,2}, Konstantin N. Borozdin\altaffilmark{1,2} 
and William C. Priedhorsky\altaffilmark{1}}

\altaffiltext{1}{Los Alamos National Laboratory, Los Alamos, NM 87545}

\altaffiltext{2}{Space Research Institute, Russian Academy of Sciences, 
Profsoyuznaya 84/32, Moscow, 117810 Russia}

\begin{abstract}

We study three transient X-ray sources, that were bright in the central 
region of M31 galaxy in the year 2000. Observations with {\em Chandra} and 
{\em XMM-Newton} allowed us for the first time in the history of X-ray 
astronomy, to build light curves of transient sources in M31 suitable for 
studying their variability on a time scale of months and, in some periods, 
weeks. The three sources demonstrate distinctly different types of X-ray 
variability and spectral evolution. XMMU~J004234.1+411808 is most likely a 
black hole candidate based on the similarity of its X-ray light curve 
and spectra to typical transient low-mass X-ray binaries observed in our 
Galaxy. The outburst of CXO~J004242.0+411608 lasted longer than a year, 
which makes the source an unusual X-ray transient. The supersoft transient 
XMMU~J004319.4+411759 is probably a classical nova-like system containing a 
magnetized, rapidly-spinning white dwarf. We estimate a total rate of X-ray 
transient outbursts in the central bulge of M31 to be of the order $\sim 10$ 
per year. The rate of the hard X-ray transients ($\sim 5$ year$^{-1}$) in 
the central part of the Andromeda Galaxy appears to be comparable to that of 
the central part of our own Galaxy.

\end{abstract}

\section{INTRODUCTION}
There are about 200 known X-ray binaries in our Galaxy. These contain either
a neutron star or a black hole (\cite{vpar95}). About one third of them
are classified as transients (see \cite{TSH96} for a review).Some of the 
X-ray transients, most notably Be-binary pulsars, erupt on regular intervals, 
while others erupt dramatically, but irregularly, with peak luminosities 
above 10$^{37}$ erg/s and typical decay timesorder a month. This last group 
represent the majority of known Galactic black hole candidates, and provides 
the best opportunity to study the processes near black hole in relative 
proximity to the Earth. Galactic transients have been therefore a focus of 
substantial observational and theoretical effort by X-ray astronomers. Much 
less, however, is known about X-ray transients in other galaxies. Previous 
observations with {\it Einstein} and {\it ROSAT} were too sparse in time for 
any systematic monitoring of even nearby galaxies. The new capabilities 
provided by {\it Chandra} and {\it XMM-Newton} started to make the difference.

X-ray observations of the Andromeda Galaxy (M31), the closest spiral galaxy 
to our own (we assume $\sim$760 kpc distance throughout this Letter - van der 
Bergh 2000), are of special interest, because M31 is in many respects similar 
to the Milky Way. M31 was observed previously with {\it Einstein} and 
{\it ROSAT}, and hundreds of X-ray sources were detected (\cite{TF91}; 
\cite{Primini93}; \cite{Supper97}; \cite{Supper01}). This Letter discusses 
transients discovered in the M31 core in recent sensitive observations by 
{\em Chandra} and {\em XMM-Newton}.

\section{OBSERVATIONS AND DATA ANALYSIS}
We analyzed a series of {\it Chandra} data consisting of seven observations 
with Advanced CCD Imaging Spectrometer (ACIS) and seven High Resolution 
Camera (HRC) observations.  We used all publicly available observations of 
M31 central region covering the period from Oct 1999 to July 2000 (Table 
\ref{obslog}). A more detailed description of the observations can be found 
in \cite{Garcia00} (hereafter G00) and \cite{DiS01}.

\begin{figure}
\epsfxsize=16cm
\epsffile{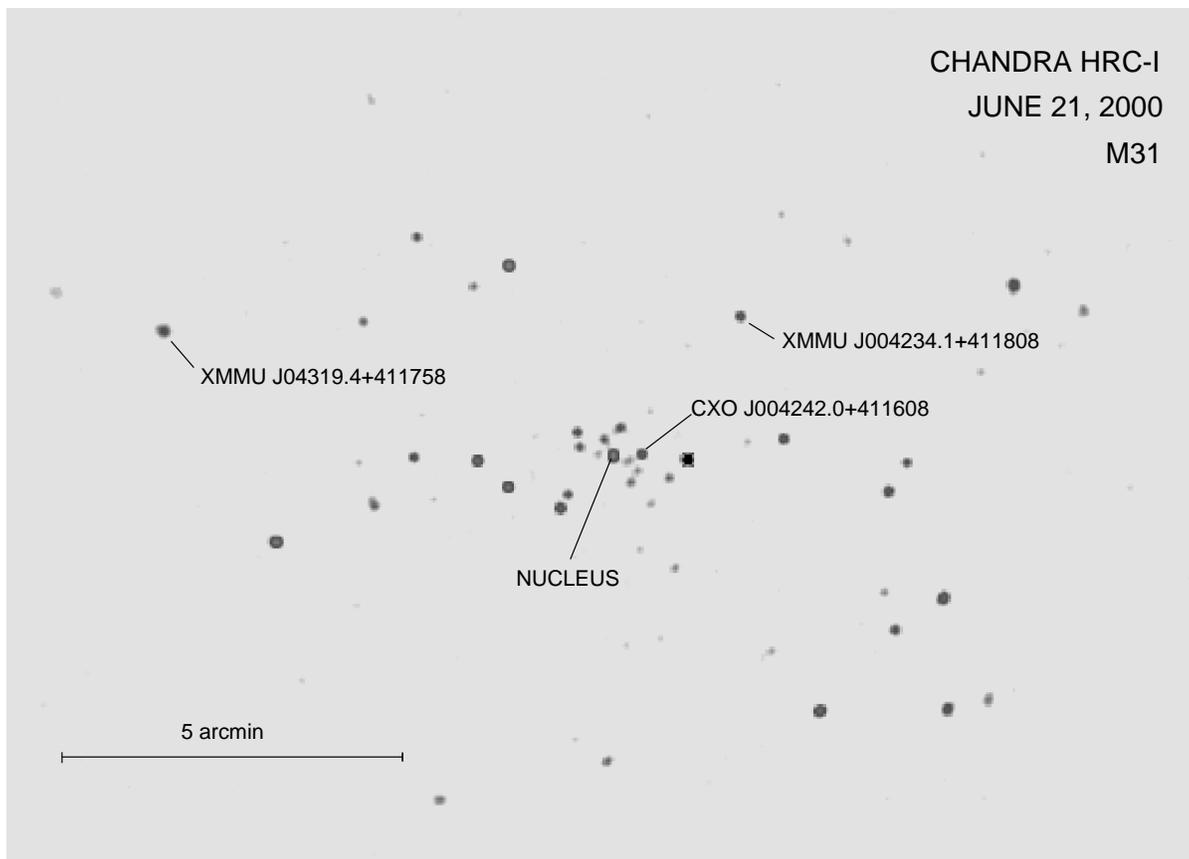}
\caption{The central region of M31 as it appears in a $1.2$ ks {\em Chandra} 
HRC-I observation (June 21, 2000). The transient sources are marked with 
arrows. The position of nuclear source is shown for comparison. 
\label{image_general}}
\end{figure}

The data were processed using the CIAO v2.1.2\footnote{http://asc.harvard.edu/ciao/} threads. We generated X-ray images of the central region of M31 
(Fig. \ref{image_general}), and performed a wavelet deconvolution of the 
images to detect and localize point sources. Because of calibration 
uncertainties below 0.3 keV and high background above 7 keV, we excuded 
ACIS data outside these energy limits. To estimate energy fluxes 
and spectra, we extracted counts within source circular regions of $\sim 3$ 
to $6\arcsec$ radii (depending on the distance of the source from the 
telescope axis). Background counts were extracted from the adjacent 
source-free regions. Spectra of the bright hard sources (XMMU~J004234.1+
411808 and CXO~J004242.0+411608) were fitted with an absorbed power law 
model using XSPEC v.11\footnote{http://heasarc.gsfc.nasa.gov/docs/xanadu/xspec/index.html}. The ACIS count rates were converted into energy fluxes using 
the analytical fits to the spectra for brighter sources or PIMMS\footnote{http://heasarc.gsfc.nasa.gov/Tools/w3pimms.html} with standard parameters 
otherwise. For the HRC the source fluxes were estimated by PIMMS, based on 
the observed count rates and the spectral parameters measured by ACIS and 
{\em XMM-Newton}.

The central region of M31 was observed with {\em XMM-Newton} on June 25 
and December 28, 2000 (\cite{Shirey01}, hereafter S01; \cite{Osborne01}, 
hereafter O01; \cite{Trudolyubov01}, hereafter T01) as a part of Performance 
Verification (PV) program (Table \ref{obslog}). In our analysis we used the 
data from two European Photon Imaging Camera (EPIC) instruments: the MOS1 
and MOS2 detectors (\cite{Turner01}). In all observations the EPIC instruments 
operated in {\em full window} mode ($30 \arcmin$ FOV) with a medium optical 
blocking filter. 
 
We reduced EPIC data with the {\em XMM-Newton} Science Analysis System (SAS 
v5.01)\footnote{http://xmm.vilspa.esa.es/user/}. To obtain the spectra of 
individual point sources, we used a source circle of $\sim 20 - 30 \arcsec$ 
and took backgrounds from adjacent source-free regions. We used data in 
the $0.3 - 7$ keV energy range, for consistency with {\it Chandra} data and 
because of uncertainties in the calibration of the EPIC instruments. This is 
the  default energy band for all fluxes and luminosities presented below.

The energy spectra of the two hard sources (XMMU~J004234.1+411808 and 
CXO~J004242.0+41608) were fitted to an absorbed power law, and an absorbed 
black body radiation model was applied to fit the spectrum of the supersoft 
transient XMMU J004319.4+411759 (Table \ref{spec_par}). Since the flux from 
the latter was negligible above 1.5 keV, we considered only its $0.3 - 1.5$ 
keV spectrum. EPIC-MOS1 and MOS2 data were fit simultaneously, but with 
independent normalizations. 

\section{RESULTS AND DISCUSSION}

\subsection{Hard X-ray Nova XMMU J004234.1+411808}
A new bright source, XMMU J004234.1+411808, was discovered on June 25, 
2000 during the first {\em XMM-Newton} observation of M31 (S01, O01, T01). 
Our analysis of the archival {\it Chandra} data revealed the presence of 
a bright X-ray source at the position consistent with XMMU J004234.1+
411808 as early as June 21, 2000, and again on July 2 and July 29 (Table 
\ref{obslog}). Based on the {\em Chandra} aspect solution, which is 
currently limited by systematics to $\sim0.6\arcsec$ accuracy, we measure 
the position of XMMU J004234.1+411808 to be $\alpha = 00^{h} 42^{m} 
34.44^{s}$, $\delta = 41\arcdeg 18\arcmin 09.7\arcsec$ (2000 equinox).
 
\begin{figure}
\epsfxsize=16cm
\epsffile{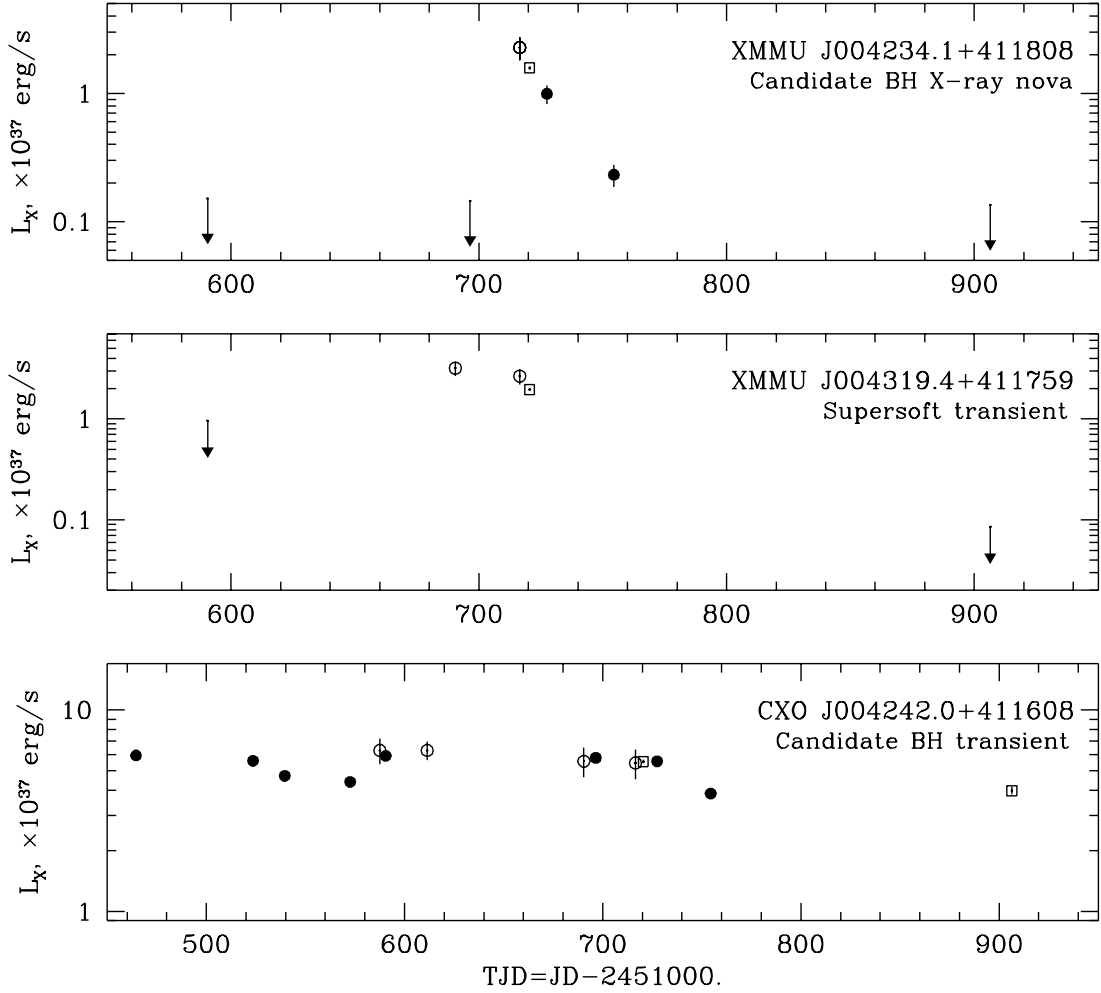}
\caption{X-ray flux histories of three transient sources in the $0.3 - 7.0$ 
keV energy range. {\em Upper panel}: hard X-ray Nova XMMU J04234.1+411808. 
{\em Middle panel}: transient supersoft source XMMU J004319.4+411758. {\em 
Lower panel}: {\em Chandra} transient CXO J004242.0+411608. {\em Filled 
circles}, {\em open circles} and {\em open squares} denote ACIS, HRC-I and 
EPIC-MOS data respectively. The X-ray luminosities are in units of $\times 
10^{37}$ erg s$^{-1}$. Upper limits correspond to a $2 \sigma$ level. 
\label{lc_general}}
\end{figure}

The X-ray light curve of XMMU J004234.1+411808  shown in Fig. 
\ref{lc_general}a. Combining the results of {\em Chandra} and {\em XMM} 
observations, we place an upper limit on the effective duration of the 
outburst of $\ltorder$40 days. We have no measurements of the outburst
rise, but it was shorter than 20 days and followed by a slower decay. 
A fit to decaying part of the light curve with an exponential gives a 
characteristic e-folding time: $\tau_{fold} = 17\pm2$ days. The source 
flux declined from $\sim 3.3 \times 10^{-13}$ erg s$^{-1}$ cm$^{-2}$ on 
June 21 to $\sim 3.4 \times 10^{-14}$ erg s$^{-1}$ cm$^{-2}$ on July 29, 
which corresponds to the luminosity drop from $\sim 2.3 \times 10^{37}$ 
erg s$^{-1}$ to $\sim 2.3 \times 10^{36}$ erg s$^{-1}$. From the ACIS 
observations prior to the outburst, and assuming the same spectrum as 
measured by {\em XMM}, we compute a $2 \sigma$ upper limit to the quiescent 
luminosity of $\sim 1.5 \times 10^{36}$ erg s$^{-1}$, $\sim 20$ times lower 
than maximum measured luminosity. 

\begin{figure}
\epsfxsize=16cm
\epsffile{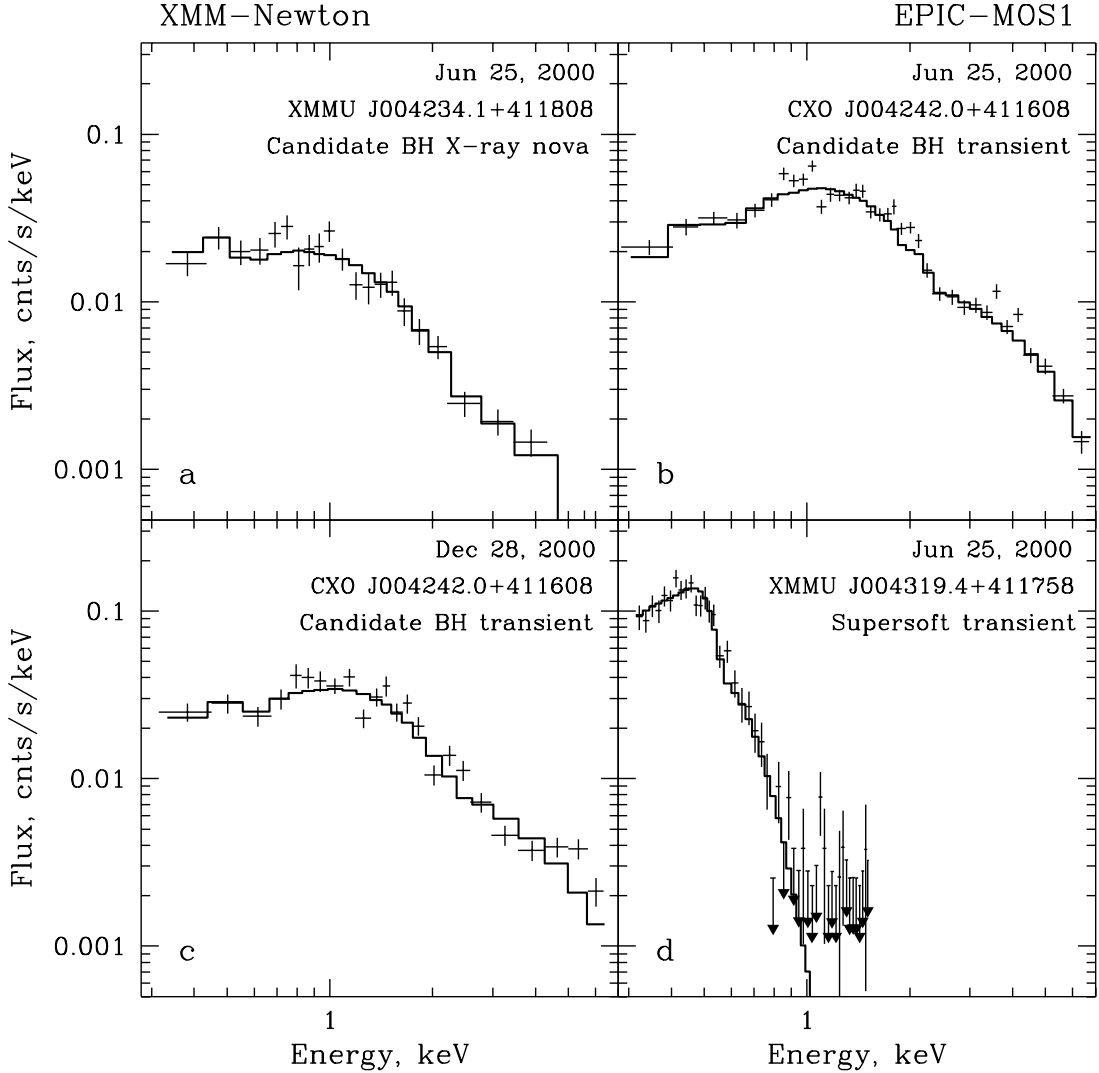}
\caption{Representative count spectra of transient sources. EPIC-MOS1 
data, $0.3 - 7$ keV energy range. Spectral model fits are shown for 
comparison. {\em Panel a}: hard X-ray Nova XMMU J04234.1+411808, Jun. 25, 
2000 observation, fit by absorbed power law model ($\alpha = 2.13$, 
N$_{\rm H} = 7 \times 10^{20}$ cm$^{-2}$). {\em Panel b}: {\em Chandra} 
transient CXO J004242.0+411608, Jun. 25, 2000 observation, fit by absorbed 
power law model ($\alpha = 1.62$, N$_{\rm H} = 1 \times 10^{21}$ cm$^{-2}$). 
{\em Panel c}: The same source, Dec. 28, 2000 observation, fit by absorbed 
power law model ($\alpha = 1.54$, N$_{\rm H} = 6 \times 10^{20}$ cm$^{-2}$). 
{\em Panel d}: transient supersoft source XMMU J004319.4+411758, Jun. 25, 
2000 observation, fit by absorbed black body radiation model (T$_{\rm bb} = 
61$ eV, N$_{\rm H} = 1.4 \times 10^{21}$ cm$^{-2}$). \label{spec_fig}}
\end{figure}

The energy spectrum of the source measured with ACIS-S on July 2 is similar
to one detected with EPIC-MOS a week earlier (Fig. \ref{spec_fig}{\em a}) 
and both may be fit to an absorbed power law with $\alpha \sim$2.1
(see Table \ref{spec_par} for fit parameters).

The energy spectrum and X-ray luminosity of XMMU J004234.1+411808 
correspond to the hard/low spectral state of Galactic black hole candidates 
(\cite{TL95}). Moreover, the X-ray evolution of the source is typical for 
well established class of X-ray novae with fast-rise-exponential-decay (FRED-
type) lightcurves (\cite{CLS97}). For many of these objects, radial velocity 
measurements proved that binary systems harbor black holes. We propose that 
XMMU J004234.1+411808 is an X-ray nova and the first stellar mass black hole 
candidate in M31.

\subsection{Transient source CXO~J004242.0+411608}
The X-ray source CXO~J004242.0+411608 was discovered on December 1999 
during the first {\em Chandra} observation of M31 (G00) and remained 
detectable throughout whole Y2K (O01, T01). The source was not detected 
with {\em XMM} on June 29, 2001 (\cite{Shirey01_1}), which sets a lower 
limit of $\sim$400 days to the outburst duration. 
       
The source exhibited remarkable stability of the X-ray luminosity 
and spectral shape throughout the outburst (Fig. \ref{lc_general}{\em b, 
c} and Table \ref{spec_par}). The average energy spectrum may be 
approximated with an absorbed power law  with $\alpha \sim 1.5 - 1.6$. 
We note that the ACIS spectra of the source tend to be generally flatter 
than EPIC-MOS spectra (Table \ref{spec_par}).  Best-fit values of 
the absorbing column are also systematically higher for the ACIS. 
Both discrepancies may be possibly attributed to the presence of a $10 - 
15 \%$ pile-up in the ACIS spectra.

Long ($\gtorder$1 year) plateaus were detected in light curves of several 
black hole candidates, whereas shorter plateaus seem more typical for 
neutron star systems (Chen et al. 1997). The hard spectrum and long plateau 
make this outburst remarkably similar to the long-lasting hard transient 
GRS 1716-249 (\cite{sun94}; \cite{rev98}).

\subsection{Supersoft transient XMMU~J004319.4+411759}

The X-ray transient source XMMU~J004319.4+\\411759 was discovered on June 
25, 2000 by {\em XMM} (S01,O01). Our analysis of May 26 {\it Chandra} data
revealed the presence of a bright X-ray source at the position $\alpha = 
00^{h} 43^{m} 19.76^{s}$, $\delta = 41\arcdeg 17\arcmin 57.6\arcsec$ 
(2000 eq., error circle $\sim 0.6\arcsec$), consistent with previously 
reported {\em XMM} position (O01).The source remained active during the 
June 21 {\em Chandra} observation. An energy spectrum obtained with EPIC-
MOS1 detector on June 25 is shown in Figure \ref{spec_fig}{\em d}. It may be 
satisfactorily fit to an absorbed black body radiation model with a 
characteristic temperature of $\sim 60$ eV and absorbing column of $\sim 
1.4 \times 10^{21}$ cm$^{-2}$. 

The spectrum of XMMU J004319.4+411759 is similar to the Galactic supersoft 
sources and may be interpreted as a result of thermonuclear burning of the 
accreted matter on the surface of the white dwarf (see \cite{KVDH97} for a 
review). The source proved to be a pulsator with a $\sim 865$s period (for 
more detailed discussion see O01). The transient behavior of the pulsator 
hints that it may be a classical Nova in the supersoft X-ray spectral phase, 
several tens of days after the peak of the outburst (\cite{KVDH97}; O01).
 
\subsection{On the rate of transient outbursts}
Given a number of transient sources detected with {\em Chandra} and 
{\em XMM}, one can estimate the total rate of transient outbursts in the 
central bulge of M31. Assuming a luminosity of $>10^{37}$ erg s$^{-1}$ 
and an average source outburst duration to be more than a month, and 
taking into account the spacing of the {\em Chandra} and {\em XMM} 
observations, we estimate a rate of $\sim 10$ year$^{-1}$. We note a large 
uncertainty of this estimate, due to the limited statistics. About 
$\sim 50\%$ of a total amount of transients detected in M31 are hard X-ray 
sources corresponding to $\sim 5$ outbursts per year. It is interesting to 
compare their number with a number of X-ray transient detected in the 
corresponding region of the Galaxy.\footnote{Very high extinction of X-rays 
towards the center of our own Galaxy prevents effective detection of 
supersoft sources with current instrumentation.} Based on regular monitoring 
of the Galactic Center with {\em RXTE} and {\em BeppoSAX}, the rate of hard 
transient outbursts in this region can be estimated as $\sim 4$ year$^{-1}$, 
which is comparable to our estimate for the central part of M31. This is 
firther evidence for the similarity between the stellar populations of the 
Galaxy and M31. 

\section{Acknowledgements}

This research has made use of data obtained through the {\em Chandra} public 
data archive. This paper is based in part on observations obtained with {\em 
XMM-Newton}, an ESA science mission with instruments and contributions 
directly funded by ESA Member States and the USA (NASA). 

The authors are grateful to Mike Watson, the Principal Investigator of the 
PV observations of M31 with {\it XMM-Newton}.

\begin{table}
\small
\caption{{\em Chandra} and {\em XMM-Newton} observations of the central 
region of M31. \label{obslog}}
\begin{tabular}{cccccc}
\hline
\hline 
   Date    &  TJD$^{a}$  &   Observatory    & Obs ID & Instrument &Exposure$^{b}$\\
   (UT)    &       &                  &        &            & (ksec)  \\             
\hline
13/10/1999 &$464$&   {\em Chandra}  &   303  &  ACIS-I    &  $9.0$\\
11/12/1999 &$523$&   {\em Chandra}  &   305  &  ACIS-I    &  $4.1$\\
27/12/1999 &$539$&   {\em Chandra}  &   306  &  ACIS-I    &  $4.1$\\
29/01/2000 &$572$&   {\em Chandra}  &   307  &  ACIS-I    &  $4.1$\\
13/02/2000 &$587$&   {\em Chandra}  &   270  &   HRC-I    &  $1.4$\\
16/02/2000 &$590$&   {\em Chandra}  &   308  &  ACIS-I    &  $4.0$\\
08/03/2000 &$611$&   {\em Chandra}  &   271  &   HRC-I    &  $2.4$\\
26/05/2000 &$690$&   {\em Chandra}  &   272  &   HRC-I    &  $1.2$\\
01/06/2000 &$696$&   {\em Chandra}  &   309  &  ACIS-S    &  $5.0$\\
21/06/2000 &$716$&   {\em Chandra}  &   273  &   HRC-I    &  $1.2$\\
25/06/2000 &$720$& {\em XMM-Newton} &   100  &    EPIC    & $34.8$\\
02/07/2000 &$727$&   {\em Chandra}  &   310  &  ACIS-S    &  $5.0$\\
29/07/2000 &$754$&   {\em Chandra}  &   311  &  ACIS-I    &  $6.0$\\
28/12/2000 &$906$& {\em XMM-Newton} &   193  &    EPIC    & $12.4$\\
\hline
\end{tabular}
\begin{list}{}{}
\item[$^{a}$] -- TJD - truncated Julian Date, TJD=JD-2451000
\item[$^{b}$] -- total exposure time
\end{list}
\end{table}

\begin{table}
\small
\caption{Best-fit model parameters of the energy spectra of transient 
sources ({\em Chandra}-ACIS and {\em XMM-Newton}-MOS data). Parameter 
errors correspond to $1 \sigma$ level.
\label{spec_par}}
\small
\begin{tabular}{ccccccc}
\hline
\hline
\multicolumn{7}{c}{Absorbed Power Law ($0.3 - 7$ keV energy range)}\\
\hline
Date & Photon &N$_{\rm H}$&Flux$^{a}$&L$_{\rm X}^{b}$&$\chi^{2}$ & Instrument\\
(UT) & Index  &($\times 10^{22}$ cm$^{-2}$)  & & &(d.o.f.)       &           \\
\hline
\multicolumn{7}{c}{CXO J004242.0+411608}\\
\hline
13/10/1999 & $1.52\pm0.10$ & $0.25\pm0.04$ & $8.63\pm0.27$ & $7.76\pm0.24$ & 58.2(44) & ACIS-I\\
11/12/1999 & $1.35\pm0.19$ & $0.13\pm0.07$ & $8.12\pm0.53$ & $6.46\pm0.42$ & 34.9(42) & ACIS-I\\
27/12/1999 & $1.60\pm0.21$ & $0.17\pm0.08$ & $6.83\pm0.43$ & $6.00\pm0.38$ & 48.6(44) & ACIS-I\\
29/01/2000 & $1.84\pm0.22$ & $0.37^{+0.10}_{-0.18}$ & $6.38\pm0.36$ & $7.03\pm0.40$ & 42.8(53) & ACIS-I\\
16/02/2000 & $1.45\pm0.17$ & $0.20\pm0.06$ & $8.59\pm0.44$ & $7.33\pm0.38$ & 52.3(62)& ACIS-I\\
01/06/2000 & $1.25\pm0.14$ & $0.11\pm0.03$ & $8.40\pm0.36$ & $6.50\pm0.28$ & 45.6(46)& ACIS-S\\
25/06/2000 & $1.61\pm0.05$ & $0.10\pm0.01$ & $8.05\pm0.16$ & $6.59\pm0.13$ & 212.2(230)& EPIC-MOS\\
02/07/2000 & $1.30\pm0.12$ & $0.10\pm0.03$ & $8.08\pm0.35$ & $6.23\pm0.27$ & 38.8(23)& ACIS-S\\
28/12/2000 & $1.54\pm0.11$ & $0.06\pm0.03$ & $5.77\pm0.25$ & $4.39\pm0.19$ & 52.3(44) & EPIC-MOS\\
\hline
\multicolumn{7}{c}{XMMU J004234.1+411808}\\
\hline
25/06/2000 & $2.13\pm0.07$ & $0.07\pm0.02$ & $2.29\pm0.07$ & $2.02\pm0.06$ & 79.9(96) & EPIC-MOS\\
02/07/2000 & $1.96\pm0.54$ & $0.07(fixed)$ & $1.44\pm0.24$ & $1.15\pm0.19$ & 4(5) & ACIS-S\\
\hline
\hline
\multicolumn{7}{c}{Absorbed Black Body ($0.3 - 1.5$ keV energy range)}\\
\hline
Date & T$_{\rm bb}$&N$_{\rm H}$&Flux$^{c}$&L$_{\rm X}^{d}$&$\chi^{2}$ & Instrument\\ 
(UT) & (eV)     &($\times 10^{22}$ cm$^{-2}$)  & & &(d.o.f.)       &              \\
\hline
\multicolumn{7}{c}{XMMU J004234.1+411808}\\
\hline
25/06/2000 & $61\pm2$ &$0.14\pm0.02$&$2.85\pm0.07$&$1.29\pm0.04$&53.8(48)& EPIC-MOS\\
\hline
\end{tabular}

\begin{list}{}{}
\item $^{a}$ -- absorbed model flux in the $0.3 - 7.0$ keV energy range in 
units of $10^{-13}$ erg s$^{-1}$ cm$^{-2}$
\item $^{b}$ -- estimated unabsorbed X-ray luminosity in the $0.3 - 7.0$ 
keV energy range in units of $10^{37}$ erg s$^{-1}$
\item $^{c}$ -- absorbed model flux in the $0.3 - 1.5$ keV energy range in 
units of $10^{-13}$ erg s$^{-1}$ cm$^{-2}$
\item $^{d}$ -- estimated unabsorbed X-ray luminosity in the $0.3 - 1.5$ keV 
energy range in units of $10^{38}$ erg s$^{-1}$
\end{list}

\end{table}

\end{document}